\documentclass{svproc}
\usepackage{xcolor}
\usepackage{geometry}
\usepackage{amsfonts}
 \usepackage{amssymb}
 \usepackage{amsmath}
 \usepackage{xspace}
\usepackage{lipsum} 
\usepackage{graphicx} 
\usepackage{subcaption} 
\usepackage{float} 
\usepackage{enumitem}
\usepackage[english]{babel}
\usepackage{cellspace}
\usepackage{hyperref}
\usepackage{multirow}
\usepackage{tikz}
\usepackage{xargs}   
\usepackage{algorithm}
\usepackage[noend]{algpseudocode}

\begin{document}
\mainmatter              
\title{Mosaic benchmark networks: Modular link streams for testing dynamic community detection algorithms}
\titlerunning{Mosaic benchmark networks}   
%

\author{Yasaman Asgari\inst{1, 3, 4} \and Remy Cazabet\inst{2} \and Pierre Borgnat\inst{1}}

\authorrunning{Y. ASGARI ET AL.}
\institute{Univ de Lyon, ENS de Lyon, CNRS, Laboratoire de Physique, F-69342 Lyon, France,\\
\and
Univ Lyon, UCBL, CNRS, INSA Lyon, LIRIS, UMR5205, F-69622 Villeurbanne, France\\
\and 
University of Zurich
UZH Digital Society Initiative
Rämisttrasse 69, 8001 Zürich\\
\and 
University of Zurich, Institut für Mathematik
Winterthurerstrasse 190, 8057 Zürich
\email{yasaman.asgari@math.uzh.ch}}
\maketitle 
\begin{abstract}
Community structure is a critical feature of real networks, providing insights into nodes' internal organization. Nowadays, with the availability of highly detailed temporal networks such as link streams, studying community structures becomes more complex due to increased data precision and time sensitivity. Despite numerous algorithms developed in the past decade for dynamic community discovery, assessing their performance on link streams remains a challenge. Synthetic benchmark graphs are a well-accepted approach for evaluating static community detection algorithms. Additionally, there have been some proposals for slowly evolving communities in low-resolution temporal networks like snapshots. Nevertheless, this approach is not yet suitable for link streams. To bridge this gap, we introduce a novel framework that generates synthetic modular link streams with predefined communities. Subsequently, we evaluate established dynamic community detection methods to uncover limitations that may not be evident in snapshots with slowly evolving communities. While no method emerges as a clear winner, we observe notable differences among them.

\end{abstract}
\section{Introduction}
Community structure is a common feature in real networks. Essentially, a community represents a network pattern where nodes have strong connections within the community and weaker connections with nodes outside\cite{girvan2002community}. Network science initially emerged when real-world temporal network data, which captures the changing structure of networks over time, was scarce. Consequently, early research on community detection primarily focused on static networks \cite{fortunato2010community}. With the increasing availability of low-resolution temporal data, such as network snapshots, attention naturally shifted to the dynamic community detection\cite{rossetti2018community}. Nowadays, we have access to highly detailed temporal networks like link streams, making the study of community structures more intricate due to the increased data granularity and time-sensitive nature.

One approach to quantitatively evaluate community detection algorithms is to employ synthetic graph generators to assess their performance and accuracy against a reference ground truth. Recently, several synthetic temporal network generators have been established to simulate snapshots with ``slowly evolving" communities, i.e., having a meaningful community structure at each discrete time step. However, link streams fail to meet this criterion. Within a link stream, the frequency of interactions per node per unit of time is exceedingly low. Consequently, during a specific time instance, we can only discern a minimal number of connections, and these connections do not reveal clear community structures. Although we can employ varying or fixed window sizes to segment our network into multiple time slices to achieve low-resolution temporal networks like snapshots, a new line of research has emerged, aiming to develop algorithms tailored to the continuous-timed nature of instantaneous edges \cite{latapy2018stream}.
Motivated by this challenge, we intend to address the issue of the absence of a benchmark network for simulating modular link streams. To address this need, we introduce the ``Mosaic benchmark network.'' Mosaic is designed to provide a reliable framework for evaluating and benchmarking dynamic community detection algorithms. It generates modular temporal networks with continuous-time edges built upon randomly planted partition networks, creating adjustable ground truths. 

The paper's organization is outlined as follows: It begins by briefly reviewing the current literature concerning benchmarks used for creating modular networks. Moving on to Section 3, a mathematical framework for link streams and the definition of a temporal community adapted for the context are presented. Then, we explain a comprehensive framework called "Mosaic" for creating link streams with communities. Finally, Section 4 involves the application of some dynamic community detection algorithms to the benchmark, aiming to uncover their capabilities and limitations.
\section{Related works}
Assessing and comparing community detection algorithms presents a significant challenge. Although real-world datasets can offer valuable insights, it has been shown that node metadata are not the same as ground truth and that treating them as such induces severe theoretical and practical problems \cite{peel2017ground}. 

To overcome this limitation, researchers have developed benchmarks to generate synthetic networks for examining algorithm behavior on networks with diverse predefined properties\cite{lancichinetti2009community}.
Synthetic network Benchmarks enable checking an algorithm against:
\begin{itemize}
    \item \textit{Definition} of communities: since there is no universal definition of community, a benchmark with its ground truth defines what we want to find and check if the method recognizes it.
    \item \textit{Stability}: the efficiency of a community detection method can be evaluated by testing it on numerous network instances that share similar characteristics. 
    \item \textit{Scalability}: gradually increasing the network size makes it possible to determine how well the algorithm handles larger and more complex networks.
\end{itemize}

Numerous network benchmarks have been introduced to establish modular static networks. Stochastic Block Models (SBM, also random planted partition graphs) \cite{holland1983stochastic} generate networks where edges between nodes in and within communities depend on a provided probability matrix.

As the need for synthetic temporal networks increased, several methods have been outlined in the literature to generate benchmark graphs for evolving communities. An evolving community scenario is defined as a structure characterized by fundamental events for communities such as birth, death, merging, splitting, expansion, contraction, iterative continuation, and the Ship of Theseus, as illustrated and described in \cite{rossetti2018community}.

Temporal network benchmarks have been developed with diverse perspectives and aims, yet they all share a common trait: they generate snapshots that reveal clear community structures. For instance, Granell et al. \cite{granell2015benchmark} propose two cyclic scenarios (migration and merge-split), and in each snapshot, communities are defined using SBMs. Bazzi et al. \cite{bazzi2016generative} introduce a method for generating multilayer networks with community structures by incorporating an SBM with additional interlayer dependency tensors. However, Cazabet et al. \cite{cazabet2020evaluating} argue that utilizing an SBM independently for generating edges in each snapshot is impractical. Therefore, they have developed a method that allows for evolving structures while maintaining the stability of most edges from one time step to the next.

Differing from all preceding approaches, our proposed benchmark introduces a framework that accomplishes two crucial objectives: 1) It enables the representation of novel scenario description generators that do not necessitate the inclusion of progressively evolving structures, and 2) It facilitates the generation of continuous-timed instantaneous edges while maintaining a low computational cost. 
\section{Mathematical Framework}
\subsection{Link stream}
Link streams, the category of temporal networks examined in this study, can be perceived as a collection of vertices denoted by \(V\), which engage with one another at specific instances, and the duration of these engagements is considered negligible. 

Based on this definition, we can formulate a link stream mathematically as:
\begin{definition}
    Link Stream: A link stream \(L\) is defined as a triple \((V, E, T)\), where \(V\) represents the set of nodes involved in interactions within a defined time domain, \(T = [T_{\text{s}}, T_{\text{e}}) \subseteq \mathbb{R}\), and \(E \subseteq V \times V \times T\) is the set of temporal edges. Each temporal edge, \(l = (u, v, t) \in E\) signifies an instantaneous interaction that took place between node \(u \in V\) and node \(v \in V\) at time \(t \in T\) \cite{latapy2018stream}.
\end{definition}

\begin{figure}[!ht]
    \centering

\tikzset{every picture/.style={line width=0.75pt}} 

\begin{tikzpicture}[x=0.75pt,y=0.75pt,yscale=-0.4,xscale=0.4]

\draw   (70.33,10) -- (600.33,10) -- (600.33,260.6) -- (70.33,260.6) -- cycle ;
\draw  [draw opacity=0][fill={rgb, 255:red, 80; green, 227; blue, 194 }  ,fill opacity=0.3 ] (70.2,10) -- (340.2,10) -- (340.2,260.2) -- (70.2,260.2) -- cycle ;
\draw  [draw opacity=0][fill={rgb, 255:red, 126; green, 211; blue, 33 }  ,fill opacity=0.3 ] (340.2,9.4) -- (601,9.4) -- (601,135.4) -- (340.2,135.4) -- cycle ;
\draw  [draw opacity=0][fill={rgb, 255:red, 189; green, 16; blue, 224 }  ,fill opacity=0.3 ] (340.2,135.4) -- (601,135.4) -- (601,260.2) -- (340.2,260.2) -- cycle ;
\draw    (71.4,280.2) -- (598.2,277.41) (125.28,275.91) -- (125.32,283.91)(179.18,275.63) -- (179.22,283.63)(233.08,275.34) -- (233.12,283.34)(286.98,275.06) -- (287.02,283.06)(340.88,274.77) -- (340.92,282.77)(394.77,274.49) -- (394.82,282.49)(448.67,274.2) -- (448.72,282.2)(502.57,273.92) -- (502.62,281.92)(556.47,273.63) -- (556.51,281.63) ;
\draw [shift={(600.2,277.4)}, rotate = 179.7] [color={rgb, 255:red, 0; green, 0; blue, 0 }  ][line width=0.75]    (10.93,-3.29) .. controls (6.95,-1.4) and (3.31,-0.3) .. (0,0) .. controls (3.31,0.3) and (6.95,1.4) .. (10.93,3.29)   ;
\draw    (60.6,10.2) -- (60.2,259.8) (64.52,60.21) -- (56.52,60.19)(64.44,110.21) -- (56.44,110.19)(64.36,160.21) -- (56.36,160.19)(64.28,210.21) -- (56.28,210.19) ;
\draw  [dash pattern={on 0.84pt off 2.51pt}]  (70.6,59.8) -- (257.8,59.8) -- (283.4,59.8) -- (599.8,59.8) ;
\draw  [dash pattern={on 0.84pt off 2.51pt}]  (71.4,109.8) -- (258.6,109.8) -- (284.2,109.8) -- (600.6,109.8) ;
\draw  [dash pattern={on 0.84pt off 2.51pt}]  (70.6,160.2) -- (257.8,160.2) -- (283.4,160.2) -- (599.8,160.2) ;
\draw  [dash pattern={on 0.84pt off 2.51pt}]  (71.8,211) -- (259,211) -- (284.6,211) -- (601,211) ;
\draw    (180,160) .. controls (171,166) and (169,204) .. (179,212) ;
\draw [shift={(179,212)}, rotate = 38.66] [color={rgb, 255:red, 0; green, 0; blue, 0 }  ][fill={rgb, 255:red, 0; green, 0; blue, 0 }  ][line width=0.75]      (0, 0) circle [x radius= 3.35, y radius= 3.35]   ;
\draw [shift={(180,160)}, rotate = 146.31] [color={rgb, 255:red, 0; green, 0; blue, 0 }  ][fill={rgb, 255:red, 0; green, 0; blue, 0 }  ][line width=0.75]      (0, 0) circle [x radius= 3.35, y radius= 3.35]   ;
\draw    (125,60) .. controls (116,66) and (114,104) .. (124,112) ;
\draw [shift={(124,112)}, rotate = 38.66] [color={rgb, 255:red, 0; green, 0; blue, 0 }  ][fill={rgb, 255:red, 0; green, 0; blue, 0 }  ][line width=0.75]      (0, 0) circle [x radius= 3.35, y radius= 3.35]   ;
\draw [shift={(125,60)}, rotate = 146.31] [color={rgb, 255:red, 0; green, 0; blue, 0 }  ][fill={rgb, 255:red, 0; green, 0; blue, 0 }  ][line width=0.75]      (0, 0) circle [x radius= 3.35, y radius= 3.35]   ;
\draw    (249,61) .. controls (240,67) and (238,105) .. (248,113) ;
\draw [shift={(248,113)}, rotate = 38.66] [color={rgb, 255:red, 0; green, 0; blue, 0 }  ][fill={rgb, 255:red, 0; green, 0; blue, 0 }  ][line width=0.75]      (0, 0) circle [x radius= 3.35, y radius= 3.35]   ;
\draw [shift={(249,61)}, rotate = 146.31] [color={rgb, 255:red, 0; green, 0; blue, 0 }  ][fill={rgb, 255:red, 0; green, 0; blue, 0 }  ][line width=0.75]      (0, 0) circle [x radius= 3.35, y radius= 3.35]   ;
\draw    (375,59) .. controls (366,65) and (364,103) .. (374,111) ;
\draw [shift={(374,111)}, rotate = 38.66] [color={rgb, 255:red, 0; green, 0; blue, 0 }  ][fill={rgb, 255:red, 0; green, 0; blue, 0 }  ][line width=0.75]      (0, 0) circle [x radius= 3.35, y radius= 3.35]   ;
\draw [shift={(375,59)}, rotate = 146.31] [color={rgb, 255:red, 0; green, 0; blue, 0 }  ][fill={rgb, 255:red, 0; green, 0; blue, 0 }  ][line width=0.75]      (0, 0) circle [x radius= 3.35, y radius= 3.35]   ;
\draw    (412,59) .. controls (403,65) and (401,103) .. (411,111) ;
\draw [shift={(411,111)}, rotate = 38.66] [color={rgb, 255:red, 0; green, 0; blue, 0 }  ][fill={rgb, 255:red, 0; green, 0; blue, 0 }  ][line width=0.75]      (0, 0) circle [x radius= 3.35, y radius= 3.35]   ;
\draw [shift={(412,59)}, rotate = 146.31] [color={rgb, 255:red, 0; green, 0; blue, 0 }  ][fill={rgb, 255:red, 0; green, 0; blue, 0 }  ][line width=0.75]      (0, 0) circle [x radius= 3.35, y radius= 3.35]   ;
\draw    (443.4,57.8) .. controls (434.4,63.8) and (432.4,101.8) .. (442.4,109.8) ;
\draw [shift={(442.4,109.8)}, rotate = 38.66] [color={rgb, 255:red, 0; green, 0; blue, 0 }  ][fill={rgb, 255:red, 0; green, 0; blue, 0 }  ][line width=0.75]      (0, 0) circle [x radius= 3.35, y radius= 3.35]   ;
\draw [shift={(443.4,57.8)}, rotate = 146.31] [color={rgb, 255:red, 0; green, 0; blue, 0 }  ][fill={rgb, 255:red, 0; green, 0; blue, 0 }  ][line width=0.75]      (0, 0) circle [x radius= 3.35, y radius= 3.35]   ;
\draw    (498,58) .. controls (489,64) and (487,102) .. (497,110) ;
\draw [shift={(497,110)}, rotate = 38.66] [color={rgb, 255:red, 0; green, 0; blue, 0 }  ][fill={rgb, 255:red, 0; green, 0; blue, 0 }  ][line width=0.75]      (0, 0) circle [x radius= 3.35, y radius= 3.35]   ;
\draw [shift={(498,58)}, rotate = 146.31] [color={rgb, 255:red, 0; green, 0; blue, 0 }  ][fill={rgb, 255:red, 0; green, 0; blue, 0 }  ][line width=0.75]      (0, 0) circle [x radius= 3.35, y radius= 3.35]   ;
\draw    (524,58) .. controls (515,64) and (513,102) .. (523,110) ;
\draw [shift={(523,110)}, rotate = 38.66] [color={rgb, 255:red, 0; green, 0; blue, 0 }  ][fill={rgb, 255:red, 0; green, 0; blue, 0 }  ][line width=0.75]      (0, 0) circle [x radius= 3.35, y radius= 3.35]   ;
\draw [shift={(524,58)}, rotate = 146.31] [color={rgb, 255:red, 0; green, 0; blue, 0 }  ][fill={rgb, 255:red, 0; green, 0; blue, 0 }  ][line width=0.75]      (0, 0) circle [x radius= 3.35, y radius= 3.35]   ;
\draw    (564,58) .. controls (555,64) and (553,102) .. (563,110) ;
\draw [shift={(563,110)}, rotate = 38.66] [color={rgb, 255:red, 0; green, 0; blue, 0 }  ][fill={rgb, 255:red, 0; green, 0; blue, 0 }  ][line width=0.75]      (0, 0) circle [x radius= 3.35, y radius= 3.35]   ;
\draw [shift={(564,58)}, rotate = 146.31] [color={rgb, 255:red, 0; green, 0; blue, 0 }  ][fill={rgb, 255:red, 0; green, 0; blue, 0 }  ][line width=0.75]      (0, 0) circle [x radius= 3.35, y radius= 3.35]   ;
\draw    (354,160) .. controls (345,166) and (343,204) .. (353,212) ;
\draw [shift={(353,212)}, rotate = 38.66] [color={rgb, 255:red, 0; green, 0; blue, 0 }  ][fill={rgb, 255:red, 0; green, 0; blue, 0 }  ][line width=0.75]      (0, 0) circle [x radius= 3.35, y radius= 3.35]   ;
\draw [shift={(354,160)}, rotate = 146.31] [color={rgb, 255:red, 0; green, 0; blue, 0 }  ][fill={rgb, 255:red, 0; green, 0; blue, 0 }  ][line width=0.75]      (0, 0) circle [x radius= 3.35, y radius= 3.35]   ;
\draw    (392,160) .. controls (383,166) and (381,204) .. (391,212) ;
\draw [shift={(391,212)}, rotate = 38.66] [color={rgb, 255:red, 0; green, 0; blue, 0 }  ][fill={rgb, 255:red, 0; green, 0; blue, 0 }  ][line width=0.75]      (0, 0) circle [x radius= 3.35, y radius= 3.35]   ;
\draw [shift={(392,160)}, rotate = 146.31] [color={rgb, 255:red, 0; green, 0; blue, 0 }  ][fill={rgb, 255:red, 0; green, 0; blue, 0 }  ][line width=0.75]      (0, 0) circle [x radius= 3.35, y radius= 3.35]   ;
\draw    (431.4,160.8) .. controls (422.4,166.8) and (420.4,204.8) .. (430.4,212.8) ;
\draw [shift={(430.4,212.8)}, rotate = 38.66] [color={rgb, 255:red, 0; green, 0; blue, 0 }  ][fill={rgb, 255:red, 0; green, 0; blue, 0 }  ][line width=0.75]      (0, 0) circle [x radius= 3.35, y radius= 3.35]   ;
\draw [shift={(431.4,160.8)}, rotate = 146.31] [color={rgb, 255:red, 0; green, 0; blue, 0 }  ][fill={rgb, 255:red, 0; green, 0; blue, 0 }  ][line width=0.75]      (0, 0) circle [x radius= 3.35, y radius= 3.35]   ;
\draw    (495,160) .. controls (486,166) and (484,204) .. (494,212) ;
\draw [shift={(494,212)}, rotate = 38.66] [color={rgb, 255:red, 0; green, 0; blue, 0 }  ][fill={rgb, 255:red, 0; green, 0; blue, 0 }  ][line width=0.75]      (0, 0) circle [x radius= 3.35, y radius= 3.35]   ;
\draw [shift={(495,160)}, rotate = 146.31] [color={rgb, 255:red, 0; green, 0; blue, 0 }  ][fill={rgb, 255:red, 0; green, 0; blue, 0 }  ][line width=0.75]      (0, 0) circle [x radius= 3.35, y radius= 3.35]   ;
\draw    (551,161) .. controls (542,167) and (540,205) .. (550,213) ;
\draw [shift={(550,213)}, rotate = 38.66] [color={rgb, 255:red, 0; green, 0; blue, 0 }  ][fill={rgb, 255:red, 0; green, 0; blue, 0 }  ][line width=0.75]      (0, 0) circle [x radius= 3.35, y radius= 3.35]   ;
\draw [shift={(551,161)}, rotate = 146.31] [color={rgb, 255:red, 0; green, 0; blue, 0 }  ][fill={rgb, 255:red, 0; green, 0; blue, 0 }  ][line width=0.75]      (0, 0) circle [x radius= 3.35, y radius= 3.35]   ;
\draw    (563,161) .. controls (554,167) and (552,205) .. (562,213) ;
\draw [shift={(562,213)}, rotate = 38.66] [color={rgb, 255:red, 0; green, 0; blue, 0 }  ][fill={rgb, 255:red, 0; green, 0; blue, 0 }  ][line width=0.75]      (0, 0) circle [x radius= 3.35, y radius= 3.35]   ;
\draw [shift={(563,161)}, rotate = 146.31] [color={rgb, 255:red, 0; green, 0; blue, 0 }  ][fill={rgb, 255:red, 0; green, 0; blue, 0 }  ][line width=0.75]      (0, 0) circle [x radius= 3.35, y radius= 3.35]   ;
\draw    (288.8,58) .. controls (267.8,77) and (266.8,187) .. (284.6,211) ;
\draw [shift={(284.6,211)}, rotate = 53.44] [color={rgb, 255:red, 0; green, 0; blue, 0 }  ][fill={rgb, 255:red, 0; green, 0; blue, 0 }  ][line width=0.75]      (0, 0) circle [x radius= 3.35, y radius= 3.35]   ;
\draw [shift={(288.8,58)}, rotate = 137.86] [color={rgb, 255:red, 0; green, 0; blue, 0 }  ][fill={rgb, 255:red, 0; green, 0; blue, 0 }  ][line width=0.75]      (0, 0) circle [x radius= 3.35, y radius= 3.35]   ;
\draw    (98.8,57) .. controls (77.8,76) and (76.8,186) .. (94.6,210) ;
\draw [shift={(94.6,210)}, rotate = 53.44] [color={rgb, 255:red, 0; green, 0; blue, 0 }  ][fill={rgb, 255:red, 0; green, 0; blue, 0 }  ][line width=0.75]      (0, 0) circle [x radius= 3.35, y radius= 3.35]   ;
\draw [shift={(98.8,57)}, rotate = 137.86] [color={rgb, 255:red, 0; green, 0; blue, 0 }  ][fill={rgb, 255:red, 0; green, 0; blue, 0 }  ][line width=0.75]      (0, 0) circle [x radius= 3.35, y radius= 3.35]   ;
\draw    (154,110) .. controls (145,116) and (143,154) .. (153,162) ;
\draw [shift={(153,162)}, rotate = 38.66] [color={rgb, 255:red, 0; green, 0; blue, 0 }  ][fill={rgb, 255:red, 0; green, 0; blue, 0 }  ][line width=0.75]      (0, 0) circle [x radius= 3.35, y radius= 3.35]   ;
\draw [shift={(154,110)}, rotate = 146.31] [color={rgb, 255:red, 0; green, 0; blue, 0 }  ][fill={rgb, 255:red, 0; green, 0; blue, 0 }  ][line width=0.75]      (0, 0) circle [x radius= 3.35, y radius= 3.35]   ;
\draw    (307,109) .. controls (292,118) and (294,198) .. (308,211) ;
\draw [shift={(308,211)}, rotate = 42.88] [color={rgb, 255:red, 0; green, 0; blue, 0 }  ][fill={rgb, 255:red, 0; green, 0; blue, 0 }  ][line width=0.75]      (0, 0) circle [x radius= 3.35, y radius= 3.35]   ;
\draw [shift={(307,109)}, rotate = 149.04] [color={rgb, 255:red, 0; green, 0; blue, 0 }  ][fill={rgb, 255:red, 0; green, 0; blue, 0 }  ][line width=0.75]      (0, 0) circle [x radius= 3.35, y radius= 3.35]   ;
\draw    (226,60) .. controls (211,69) and (212,145) .. (226,158) ;
\draw [shift={(226,158)}, rotate = 42.88] [color={rgb, 255:red, 0; green, 0; blue, 0 }  ][fill={rgb, 255:red, 0; green, 0; blue, 0 }  ][line width=0.75]      (0, 0) circle [x radius= 3.35, y radius= 3.35]   ;
\draw [shift={(226,60)}, rotate = 149.04] [color={rgb, 255:red, 0; green, 0; blue, 0 }  ][fill={rgb, 255:red, 0; green, 0; blue, 0 }  ][line width=0.75]      (0, 0) circle [x radius= 3.35, y radius= 3.35]   ;

\draw (324.1,291.4) node [anchor=north west][inner sep=0.75pt]    {${\displaystyle T}$};
\draw (0,130.13) node [anchor=north west][inner sep=0.75pt]    {$V$};
\draw (25,50) node [anchor=north west][inner sep=0.75pt]   [align=left] {$\displaystyle v_{1}$};
\draw (25,100) node [anchor=north west][inner sep=0.75pt]   [align=left] {$\displaystyle v_{2}$};
\draw (25,150) node [anchor=north west][inner sep=0.75pt]   [align=left] {$\displaystyle v_{3}$};
\draw (25,200) node [anchor=north west][inner sep=0.75pt]   [align=left] {$\displaystyle v_{4}$};
\draw (165.5,115) node [anchor=north west][inner sep=0.75pt]  [font=\Large]  {$c_{1}$};
\draw (440.5,70.13) node [anchor=north west][inner sep=0.75pt]  [font=\Large]  {$c_{2}$};
\draw (435.1,169.73) node [anchor=north west][inner sep=0.75pt]  [font=\Large]  {$c_{3}$};

\end{tikzpicture}

    \caption{ \textbf{Modular Link stream}: A link stream \((V, E, T)\) is shown, containing 4 nodes \(V=\{v_1, v_2,v_3,v_4\}\) interacting several times with each other within the time domain \(T=[0,10)\). A mosaic partitioning, \(\mathcal{C}=\{c_1, c_2,c_3\}\) is also observed. This partitioning covers \(\{v_1, v_2, v_3,v_4\}\times [0,10)\) without any overlap.}
    \label{fig: linkstream}
\end{figure}
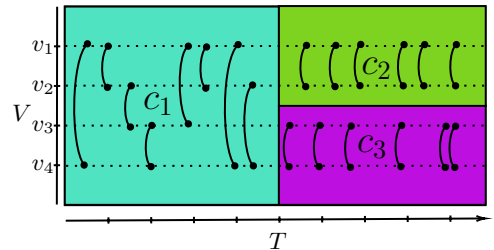

The illustration in Fig. \ref{fig: linkstream} presents a link stream featuring a set of vertices \(V=\{v_1, v_2, v_3, v_4\}\) where multiple temporal edges are observed. For example, nodes \(v_1\) and \(v_4\) establish a connection twice within the given time domain, \(T=[0,10)\).

\subsection{Mosaic: A definition for a community in link streams}
Defining a community in a link stream is a challenging task due to the fine-resolution temporal dimension involved. However, any new definition must align with the intuitive understanding of real-world applications. To give an intuition of the meaning of communities in this setting to unfamiliar readers, we illustrate it using a well-known link-stream dataset.

Sociopatterns\footnote[1]{\url{http://www.sociopatterns.org/}}  is a renowned database for real-world link streams acquired in various contexts since 2008 \cite{10.1371/journal.pone.0011596}. In these experiments, RFID sensors track real-time proximity, creating co-presence networks between individuals. For example, a substantial dataset comes from a primary school study where 230 pupils and 10 teachers wore sensors for two consecutive days. This study recorded 125,000 face-to-face interactions over 32 hours, with a temporal resolution of 20 seconds. Previous research often viewed its communities as evolving structures influenced by node and edge additions/deletions, identifying them by segmenting time into different slices with various window sizes and then applying dynamic community detection techniques.

In this study, we can observe communities emerging during specific timeframes from students interacting. For instance, students and teachers interact during lecture hours within their respective classes. During lunchtime, students with stronger friendships tend to dine together. An empty community exists at night, indicating no interactions occur during that period. 

This perspective can be extended to other contexts like people discussing a particular topic on social networks, company meetings, or sports players participating in a match together. Building on this idea, we introduce a new definition called ``Mosaic,'' which is a straightforward adaptation of non-overlapping communities from static networks to link streams.

A ``Mosaic'' community is defined as follows:
\begin{definition}
A Mosaic, denoted as \(c\), is defined as a pair of (nodes, period): \(c=(V_c, T_c)\). \(V_c\) is set of \(n\) nodes denoted as \(\{v_1,v_2,\cdots,v_n\}\). \(T_c \subset \mathbb{R}\) is an time interval, \(T_c=[T_{cs}, T_{ce})\) where, \(T_{cs}\) and \(T_{ce}\) represent the start and end times of a Mosaic \(c\), respectively. It represents the interval in which nodes \(V\) are considered part of the community \(c\).
\end{definition}

According to this definition, each node is assigned to only one community at any given time, and these communities collectively cover the entire link stream; refer to Fig. \ref{fig: linkstream} for an example. In cases where nodes do not interact significantly for a certain duration, they are assigned to an ``empty community," \(c_*\).
Therefore, a Mosaic partitioning can be defined as follows:
\begin{definition}
Mosaic partitioning: Given a link stream \(L=(V, E, T)\), \(\mathcal{C}\) is a partitioning containing \(k\) mosaics, \(\{c_1,c_2,\cdots, c_k, c_*\}\), that cover the link stream fully without any overlap. This requirement can be written as follows:
\[ \bigcup_{c\in \mathcal{C}} {V_{c}\times T_{c}}=V_L\times T_L \;\;\;\;\;\;\;\;\;\; \bigcap_{c\in \mathcal{C}} {V_{c}\times T_{c}}=\varnothing 
\]
The mosaic \(c_*\) stands for an empty community.
\end{definition}

\subsection{Mosaic Link Stream Benchmark}
\begin{figure}[!ht]
    \centering

\tikzset{every picture/.style={line width=0.75pt}} 

\begin{tikzpicture}[x=0.75pt,y=0.75pt,yscale=-0.8,xscale=0.7]

\draw   (27.81,39.65) -- (192.13,39.65) -- (192.13,141.99) -- (27.81,141.99) -- cycle ;
\draw  [draw opacity=0][fill={rgb, 255:red, 80; green, 227; blue, 194 }  ,fill opacity=0.3 ] (27.77,39.65) -- (111.48,39.65) -- (111.48,108) -- (27.77,108) -- cycle ;
\draw  [draw opacity=0][fill={rgb, 255:red, 126; green, 211; blue, 33 }  ,fill opacity=0.3 ] (111.48,39.4) -- (141.33,39.4) -- (141.33,90.86) -- (111.48,90.86) -- cycle ;
\draw  [draw opacity=0][fill={rgb, 255:red, 189; green, 16; blue, 224 }  ,fill opacity=0.3 ] (111.48,90.86) -- (192.33,90.86) -- (192.33,141.83) -- (111.48,141.83) -- cycle ;
\draw    (28.14,150) -- (190.09,148.87) (82.01,145.62) -- (82.07,153.62)(135.91,145.25) -- (135.97,153.25)(189.81,144.87) -- (189.87,152.87) ;
\draw [shift={(192.09,148.86)}, rotate = 179.6] [color={rgb, 255:red, 0; green, 0; blue, 0 }  ][line width=0.75]    (10.93,-3.29) .. controls (6.95,-1.4) and (3.31,-0.3) .. (0,0) .. controls (3.31,0.3) and (6.95,1.4) .. (10.93,3.29)   ;
\draw    (21.79,43.73) -- (21.33,142) (25.56,93.74) -- (17.56,93.71) ;
\draw  [draw opacity=0][fill={rgb, 255:red, 245; green, 166; blue, 35 }  ,fill opacity=0.3 ] (141.33,39.4) -- (191.33,39.4) -- (191.33,90.86) -- (141.33,90.86) -- cycle ;
\draw  [draw opacity=0][fill={rgb, 255:red, 74; green, 144; blue, 226 }  ,fill opacity=0.4 ] (27.77,108) -- (111.48,108) -- (111.48,142) -- (27.77,142) -- cycle ;
\draw   (29.93,212.65) -- (194.25,212.65) -- (194.25,314.99) -- (29.93,314.99) -- cycle ;
\draw  [draw opacity=0][fill={rgb, 255:red, 80; green, 227; blue, 194 }  ,fill opacity=0.3 ] (29.89,212.65) -- (113.6,212.65) -- (113.6,281) -- (29.89,281) -- cycle ;
\draw  [draw opacity=0][fill={rgb, 255:red, 126; green, 211; blue, 33 }  ,fill opacity=0.3 ] (113.6,212.4) -- (143.45,212.4) -- (143.45,263.86) -- (113.6,263.86) -- cycle ;
\draw  [draw opacity=0][fill={rgb, 255:red, 189; green, 16; blue, 224 }  ,fill opacity=0.3 ] (113.6,263.86) -- (194.45,263.86) -- (194.45,314.83) -- (113.6,314.83) -- cycle ;
\draw    (30.26,323) -- (192.21,321.87) (84.13,318.62) -- (84.19,326.62)(138.03,318.25) -- (138.09,326.25)(191.93,317.87) -- (191.99,325.87) ;
\draw [shift={(194.21,321.86)}, rotate = 179.6] [color={rgb, 255:red, 0; green, 0; blue, 0 }  ][line width=0.75]    (10.93,-3.29) .. controls (6.95,-1.4) and (3.31,-0.3) .. (0,0) .. controls (3.31,0.3) and (6.95,1.4) .. (10.93,3.29)   ;
\draw    (23.91,216.73) -- (23.45,315) (27.68,266.74) -- (19.68,266.71) ;
\draw  [draw opacity=0][fill={rgb, 255:red, 155; green, 155; blue, 155 }  ,fill opacity=0.5 ] (143.45,212.4) -- (193.45,212.4) -- (193.45,263.86) -- (143.45,263.86) -- cycle ;
\draw  [draw opacity=0][fill={rgb, 255:red, 155; green, 155; blue, 155 }  ,fill opacity=0.5 ] (29.89,281) -- (113.6,281) -- (113.6,315) -- (29.89,315) -- cycle ;
\draw   (509.99,45.7) -- (654.33,45.7) -- (654.33,143.93) -- (509.99,143.93) -- cycle ;
\draw  [draw opacity=0][fill={rgb, 255:red, 80; green, 227; blue, 194 }  ,fill opacity=0.3 ] (509.99,45.7) -- (583.52,45.7) -- (583.52,111.3) -- (509.99,111.3) -- cycle ;
\draw    (510.28,151.61) -- (652.3,150.53) (564.15,147.2) -- (564.21,155.2)(618.05,146.79) -- (618.11,154.79) ;
\draw [shift={(654.3,150.51)}, rotate = 179.56] [color={rgb, 255:red, 0; green, 0; blue, 0 }  ][line width=0.75]    (10.93,-3.29) .. controls (6.95,-1.4) and (3.31,-0.3) .. (0,0) .. controls (3.31,0.3) and (6.95,1.4) .. (10.93,3.29)   ;
\draw  [fill={rgb, 255:red, 80; green, 227; blue, 194 }  ,fill opacity=0.3 ] (220.33,117.83) .. controls (220.33,114.61) and (222.95,112) .. (226.17,112) .. controls (229.39,112) and (232,114.61) .. (232,117.83) .. controls (232,121.05) and (229.39,123.67) .. (226.17,123.67) .. controls (222.95,123.67) and (220.33,121.05) .. (220.33,117.83) -- cycle ;
\draw  [fill={rgb, 255:red, 80; green, 227; blue, 194 }  ,fill opacity=0.3 ] (297.67,112.33) .. controls (297.67,109.11) and (300.28,106.5) .. (303.5,106.5) .. controls (306.72,106.5) and (309.33,109.11) .. (309.33,112.33) .. controls (309.33,115.55) and (306.72,118.17) .. (303.5,118.17) .. controls (300.28,118.17) and (297.67,115.55) .. (297.67,112.33) -- cycle ;
\draw  [fill={rgb, 255:red, 80; green, 227; blue, 194 }  ,fill opacity=0.3 ] (257.33,146.83) .. controls (257.33,143.61) and (259.95,141) .. (263.17,141) .. controls (266.39,141) and (269,143.61) .. (269,146.83) .. controls (269,150.05) and (266.39,152.67) .. (263.17,152.67) .. controls (259.95,152.67) and (257.33,150.05) .. (257.33,146.83) -- cycle ;
\draw  [fill={rgb, 255:red, 80; green, 227; blue, 194 }  ,fill opacity=0.3 ] (233.33,74.83) .. controls (233.33,71.61) and (235.95,69) .. (239.17,69) .. controls (242.39,69) and (245,71.61) .. (245,74.83) .. controls (245,78.05) and (242.39,80.67) .. (239.17,80.67) .. controls (235.95,80.67) and (233.33,78.05) .. (233.33,74.83) -- cycle ;
\draw    (239.17,80.67) -- (232,117.83) ;
\draw    (297.67,112.33) -- (275.33,76.67) ;
\draw    (232,117.83) -- (297.67,112.33) ;
\draw    (263.17,141) -- (297.67,112.33) ;
\draw  [fill={rgb, 255:red, 80; green, 227; blue, 194 }  ,fill opacity=0.3 ] (269.5,70.83) .. controls (269.5,67.61) and (272.11,65) .. (275.33,65) .. controls (278.56,65) and (281.17,67.61) .. (281.17,70.83) .. controls (281.17,74.05) and (278.56,76.67) .. (275.33,76.67) .. controls (272.11,76.67) and (269.5,74.05) .. (269.5,70.83) -- cycle ;
\draw   (511.02,223.63) -- (654.15,223.63) -- (654.15,320.87) -- (511.02,320.87) -- cycle ;
\draw  [draw opacity=0][fill={rgb, 255:red, 126; green, 211; blue, 33 }  ,fill opacity=0.3 ] (583.9,223.4) -- (609.91,223.4) -- (609.91,272.29) -- (583.9,272.29) -- cycle ;
\draw  [draw opacity=0][fill={rgb, 255:red, 189; green, 16; blue, 224 }  ,fill opacity=0.3 ] (583.9,272.29) -- (654.33,272.29) -- (654.33,320.71) -- (583.9,320.71) -- cycle ;
\draw    (511.31,328.48) -- (652.12,327.4) (565.18,324.07) -- (565.24,332.07)(619.08,323.66) -- (619.14,331.65) ;
\draw [shift={(654.12,327.39)}, rotate = 179.56] [color={rgb, 255:red, 0; green, 0; blue, 0 }  ][line width=0.75]    (10.93,-3.29) .. controls (6.95,-1.4) and (3.31,-0.3) .. (0,0) .. controls (3.31,0.3) and (6.95,1.4) .. (10.93,3.29)   ;
\draw    (505.78,227.51) -- (505.38,320.87) (509.57,277.53) -- (501.57,277.49) ;
\draw  [fill={rgb, 255:red, 189; green, 16; blue, 224 }  ,fill opacity=0.3 ] (224.33,311.83) .. controls (224.33,308.61) and (226.95,306) .. (230.17,306) .. controls (233.39,306) and (236,308.61) .. (236,311.83) .. controls (236,315.05) and (233.39,317.67) .. (230.17,317.67) .. controls (226.95,317.67) and (224.33,315.05) .. (224.33,311.83) -- cycle ;
\draw  [fill={rgb, 255:red, 189; green, 16; blue, 224 }  ,fill opacity=0.3 ] (253.33,311.83) .. controls (253.33,308.61) and (255.95,306) .. (259.17,306) .. controls (262.39,306) and (265,308.61) .. (265,311.83) .. controls (265,315.05) and (262.39,317.67) .. (259.17,317.67) .. controls (255.95,317.67) and (253.33,315.05) .. (253.33,311.83) -- cycle ;
\draw  [fill={rgb, 255:red, 189; green, 16; blue, 224 }  ,fill opacity=0.3 ] (279.33,311.83) .. controls (279.33,308.61) and (281.95,306) .. (285.17,306) .. controls (288.39,306) and (291,308.61) .. (291,311.83) .. controls (291,315.05) and (288.39,317.67) .. (285.17,317.67) .. controls (281.95,317.67) and (279.33,315.05) .. (279.33,311.83) -- cycle ;
\draw  [fill={rgb, 255:red, 189; green, 16; blue, 224 }  ,fill opacity=0.3 ] (304.33,311.83) .. controls (304.33,308.61) and (306.95,306) .. (310.17,306) .. controls (313.39,306) and (316,308.61) .. (316,311.83) .. controls (316,315.05) and (313.39,317.67) .. (310.17,317.67) .. controls (306.95,317.67) and (304.33,315.05) .. (304.33,311.83) -- cycle ;
\draw  [fill={rgb, 255:red, 126; green, 211; blue, 33 }  ,fill opacity=0.3 ] (241.33,241.83) .. controls (241.33,238.61) and (243.95,236) .. (247.17,236) .. controls (250.39,236) and (253,238.61) .. (253,241.83) .. controls (253,245.05) and (250.39,247.67) .. (247.17,247.67) .. controls (243.95,247.67) and (241.33,245.05) .. (241.33,241.83) -- cycle ;
\draw  [fill={rgb, 255:red, 126; green, 211; blue, 33 }  ,fill opacity=0.3 ] (262.33,241.83) .. controls (262.33,238.61) and (264.95,236) .. (268.17,236) .. controls (271.39,236) and (274,238.61) .. (274,241.83) .. controls (274,245.05) and (271.39,247.67) .. (268.17,247.67) .. controls (264.95,247.67) and (262.33,245.05) .. (262.33,241.83) -- cycle ;
\draw  [fill={rgb, 255:red, 126; green, 211; blue, 33 }  ,fill opacity=0.3 ] (281.33,241.83) .. controls (281.33,238.61) and (283.95,236) .. (287.17,236) .. controls (290.39,236) and (293,238.61) .. (293,241.83) .. controls (293,245.05) and (290.39,247.67) .. (287.17,247.67) .. controls (283.95,247.67) and (281.33,245.05) .. (281.33,241.83) -- cycle ;
\draw    (230.17,306) -- (247.17,247.67) ;
\draw    (259.17,306) -- (247.17,247.67) ;
\draw    (285.17,306) -- (268.17,247.67) ;
\draw    (310.17,306) -- (287.17,247.67) ;
\draw    (285.17,306) -- (287.17,247.67) ;
\draw    (343.33,110) -- (463.33,110) ;
\draw [shift={(463.33,110)}, rotate = 180] [color={rgb, 255:red, 0; green, 0; blue, 0 }  ][line width=0.75]    (0,5.59) -- (0,-5.59)   ;
\draw [shift={(343.33,110)}, rotate = 180] [color={rgb, 255:red, 0; green, 0; blue, 0 }  ][line width=0.75]    (0,5.59) -- (0,-5.59)   ;
\draw  [color={rgb, 255:red, 0; green, 0; blue, 0 }  ,draw opacity=1 ][fill={rgb, 255:red, 0; green, 0; blue, 0 }  ,fill opacity=1 ] (353,110) .. controls (353,108.34) and (354.34,107) .. (356,107) .. controls (357.66,107) and (359,108.34) .. (359,110) .. controls (359,111.66) and (357.66,113) .. (356,113) .. controls (354.34,113) and (353,111.66) .. (353,110) -- cycle ;
\draw  [color={rgb, 255:red, 0; green, 0; blue, 0 }  ,draw opacity=1 ][fill={rgb, 255:red, 0; green, 0; blue, 0 }  ,fill opacity=1 ] (367,110) .. controls (367,108.34) and (368.34,107) .. (370,107) .. controls (371.66,107) and (373,108.34) .. (373,110) .. controls (373,111.66) and (371.66,113) .. (370,113) .. controls (368.34,113) and (367,111.66) .. (367,110) -- cycle ;
\draw  [color={rgb, 255:red, 0; green, 0; blue, 0 }  ,draw opacity=1 ][fill={rgb, 255:red, 0; green, 0; blue, 0 }  ,fill opacity=1 ] (403.33,110) .. controls (403.33,108.34) and (404.68,107) .. (406.33,107) .. controls (407.99,107) and (409.33,108.34) .. (409.33,110) .. controls (409.33,111.66) and (407.99,113) .. (406.33,113) .. controls (404.68,113) and (403.33,111.66) .. (403.33,110) -- cycle ;
\draw  [color={rgb, 255:red, 0; green, 0; blue, 0 }  ,draw opacity=1 ][fill={rgb, 255:red, 0; green, 0; blue, 0 }  ,fill opacity=1 ] (437,110) .. controls (437,108.34) and (438.34,107) .. (440,107) .. controls (441.66,107) and (443,108.34) .. (443,110) .. controls (443,111.66) and (441.66,113) .. (440,113) .. controls (438.34,113) and (437,111.66) .. (437,110) -- cycle ;
\draw  [dash pattern={on 4.5pt off 4.5pt}]  (289.33,87) .. controls (312.85,65.44) and (366.15,38.12) .. (385.21,55.86) ;
\draw [shift={(386.33,57)}, rotate = 228.01] [color={rgb, 255:red, 0; green, 0; blue, 0 }  ][line width=0.75]    (10.93,-3.29) .. controls (6.95,-1.4) and (3.31,-0.3) .. (0,0) .. controls (3.31,0.3) and (6.95,1.4) .. (10.93,3.29)   ;
\draw  [dash pattern={on 0.84pt off 2.51pt}]  (510.34,62.3) -- (586.22,61.81) ;
\draw  [dash pattern={on 0.84pt off 2.51pt}]  (511.22,100.69) -- (587.38,101.02) ;
\draw    (516.75,62.29) .. controls (521.96,65.75) and (522.25,95.16) .. (516.47,99.77) ;
\draw    (521.96,62.29) .. controls (527.16,65.75) and (527.45,95.16) .. (521.67,99.77) ;
\draw    (556.06,62.87) .. controls (561.27,66.33) and (561.56,95.74) .. (555.77,100.35) ;
\draw    (574.85,62.87) .. controls (580.05,66.33) and (580.34,95.74) .. (574.56,100.35) ;
\draw    (336.33,282.84) -- (467.77,282.84) ;
\draw [shift={(467.77,282.84)}, rotate = 180] [color={rgb, 255:red, 0; green, 0; blue, 0 }  ][line width=0.75]    (0,5.59) -- (0,-5.59)   ;
\draw [shift={(336.33,282.84)}, rotate = 180] [color={rgb, 255:red, 0; green, 0; blue, 0 }  ][line width=0.75]    (0,5.59) -- (0,-5.59)   ;
\draw  [color={rgb, 255:red, 0; green, 0; blue, 0 }  ,draw opacity=1 ][fill={rgb, 255:red, 0; green, 0; blue, 0 }  ,fill opacity=1 ] (346.11,282.84) .. controls (346.11,281.12) and (347.58,279.73) .. (349.39,279.73) .. controls (351.21,279.73) and (352.68,281.12) .. (352.68,282.84) .. controls (352.68,284.56) and (351.21,285.96) .. (349.39,285.96) .. controls (347.58,285.96) and (346.11,284.56) .. (346.11,282.84) -- cycle ;
\draw  [color={rgb, 255:red, 0; green, 0; blue, 0 }  ,draw opacity=1 ][fill={rgb, 255:red, 0; green, 0; blue, 0 }  ,fill opacity=1 ] (427.57,282.84) .. controls (427.57,281.12) and (429.04,279.73) .. (430.85,279.73) .. controls (432.67,279.73) and (434.14,281.12) .. (434.14,282.84) .. controls (434.14,284.56) and (432.67,285.96) .. (430.85,285.96) .. controls (429.04,285.96) and (427.57,284.56) .. (427.57,282.84) -- cycle ;
\draw  [color={rgb, 255:red, 0; green, 0; blue, 0 }  ,draw opacity=1 ][fill={rgb, 255:red, 0; green, 0; blue, 0 }  ,fill opacity=1 ] (412.92,282.84) .. controls (412.92,281.12) and (414.39,279.73) .. (416.21,279.73) .. controls (418.02,279.73) and (419.5,281.12) .. (419.5,282.84) .. controls (419.5,284.56) and (418.02,285.96) .. (416.21,285.96) .. controls (414.39,285.96) and (412.92,284.56) .. (412.92,282.84) -- cycle ;
\draw    (587.79,255.51) .. controls (592.83,258.82) and (593.11,287.04) .. (587.51,291.46) ;
\draw    (604.28,256.46) .. controls (609.32,259.78) and (609.6,288) .. (604,292.42) ;
\draw    (598.92,232.37) .. controls (603.97,235.69) and (604.53,298.94) .. (598.92,303.36) ;
\draw    (496.78,48.51) -- (496.38,141.87) (500.57,98.53) -- (492.57,98.49) ;
\draw   (255.93,383.65) -- (420.25,383.65) -- (420.25,485.99) -- (255.93,485.99) -- cycle ;
\draw  [draw opacity=0][fill={rgb, 255:red, 80; green, 227; blue, 194 }  ,fill opacity=0.3 ] (255.89,383.65) -- (339.6,383.65) -- (339.6,452) -- (255.89,452) -- cycle ;
\draw  [draw opacity=0][fill={rgb, 255:red, 126; green, 211; blue, 33 }  ,fill opacity=0.3 ] (339.6,383.4) -- (369.45,383.4) -- (369.45,434.86) -- (339.6,434.86) -- cycle ;
\draw  [draw opacity=0][fill={rgb, 255:red, 189; green, 16; blue, 224 }  ,fill opacity=0.3 ] (339.6,434.86) -- (420.45,434.86) -- (420.45,485.83) -- (339.6,485.83) -- cycle ;
\draw    (256.26,494) -- (418.21,492.87) (310.13,489.62) -- (310.19,497.62)(364.03,489.25) -- (364.09,497.25)(417.93,488.87) -- (417.99,496.87) ;
\draw [shift={(420.21,492.86)}, rotate = 179.6] [color={rgb, 255:red, 0; green, 0; blue, 0 }  ][line width=0.75]    (10.93,-3.29) .. controls (6.95,-1.4) and (3.31,-0.3) .. (0,0) .. controls (3.31,0.3) and (6.95,1.4) .. (10.93,3.29)   ;
\draw    (249.91,387.73) -- (249.45,486) (253.68,437.74) -- (245.68,437.71) ;
\draw  [draw opacity=0][fill={rgb, 255:red, 155; green, 155; blue, 155 }  ,fill opacity=0.5 ] (369.45,383.4) -- (419.45,383.4) -- (419.45,434.86) -- (369.45,434.86) -- cycle ;
\draw  [draw opacity=0][fill={rgb, 255:red, 155; green, 155; blue, 155 }  ,fill opacity=0.5 ] (255.89,452) -- (339.6,452) -- (339.6,486) -- (255.89,486) -- cycle ;
\draw [line width=1.5]  [dash pattern={on 5.63pt off 4.5pt}]  (395.06,440.87) .. controls (400.27,444.33) and (400.56,473.74) .. (394.77,478.35) ;
\draw [line width=1.5]    (343.33,392) .. controls (348.54,395.46) and (349.11,476.39) .. (343.33,481) ;
\draw [color={rgb, 255:red, 255; green, 0; blue, 0 }  ,draw opacity=1 ] [dash pattern={on 4.5pt off 4.5pt}]  (398.33,458) .. controls (391.44,455.05) and (386.48,442.39) .. (348.53,448.68) ;
\draw [shift={(346.77,448.98)}, rotate = 350] [color={rgb, 255:red, 255; green, 0; blue, 0 }  ,draw opacity=1 ][line width=0.75]    (10.93,-3.29) .. controls (6.95,-1.4) and (3.31,-0.3) .. (0,0) .. controls (3.31,0.3) and (6.95,1.4) .. (10.93,3.29)   ;
\draw    (263.17,141) -- (232,117.83) ;
\draw    (269.5,70.83) -- (232,117.83) ;
\draw  [dash pattern={on 4.5pt off 4.5pt}]  (323.33,266.33) .. controls (346.73,244.88) and (382.49,215.51) .. (400.95,232.59) ;
\draw [shift={(402.33,234)}, rotate = 228.01] [color={rgb, 255:red, 0; green, 0; blue, 0 }  ][line width=0.75]    (10.93,-3.29) .. controls (6.95,-1.4) and (3.31,-0.3) .. (0,0) .. controls (3.31,0.3) and (6.95,1.4) .. (10.93,3.29)   ;

\draw (96.1,164.4) node [anchor=north west][inner sep=0.75pt]    {${\displaystyle T}$};
\draw (2.38,91.17) node [anchor=north west][inner sep=0.75pt]    {$V$};
\draw (60.86,67.33) node [anchor=north west][inner sep=0.75pt]  [font=\normalsize]  {$c_{1}$};
\draw (116.94,57.89) node [anchor=north west][inner sep=0.75pt]  [font=\normalsize]  {$c_{2}$};
\draw (16.5,16) node [anchor=north west][inner sep=0.75pt]   [align=left] {\textbf{A. Scenario generation}};
\draw (152.94,59.89) node [anchor=north west][inner sep=0.75pt]  [font=\normalsize]  {$c_{4}$};
\draw (60.94,117.89) node [anchor=north west][inner sep=0.75pt]  [font=\normalsize]  {$c_{5}$};
\draw (98.22,337.4) node [anchor=north west][inner sep=0.75pt]    {${\displaystyle T}$};
\draw (4.5,264.17) node [anchor=north west][inner sep=0.75pt]    {$V$};
\draw (62.98,240.33) node [anchor=north west][inner sep=0.75pt]  [font=\normalsize]  {$c_{1}$};
\draw (119.06,230.89) node [anchor=north west][inner sep=0.75pt]  [font=\normalsize]  {$c_{2}$};
\draw (141.32,285.65) node [anchor=north west][inner sep=0.75pt]  [font=\normalsize]  {$c_{3}$};
\draw (162.06,233.89) node [anchor=north west][inner sep=0.75pt]  [font=\normalsize]  {$c_{*}$};
\draw (23.5,185) node [anchor=north west][inner sep=0.75pt]   [align=left] {\textbf{B. Empty Mosaics (}$\displaystyle \gamma $\textbf{)}};
\draw (63.06,292.89) node [anchor=north west][inner sep=0.75pt]  [font=\normalsize]  {$c_{*}$};
\draw (578.32,161.87) node [anchor=north west][inner sep=0.75pt]    {${\displaystyle T}$};
\draw (537.86,72.12) node [anchor=north west][inner sep=0.75pt]  [font=\normalsize]  {$c_{1}$};
\draw (307.5,16) node [anchor=north west][inner sep=0.75pt]   [align=left] {\textbf{C. Internal edges generation}};
\draw (231.5,44.4) node [anchor=north west][inner sep=0.75pt]    {$\alpha \ \rightarrow p_{i}{}_{n}$};
\draw (304.5,193) node [anchor=north west][inner sep=0.75pt]   [align=left] {\textbf{D. External edges generation}};
\draw (569.54,331.47) node [anchor=north west][inner sep=0.75pt]    {${\displaystyle T}$};
\draw (487.78,272.45) node [anchor=north west][inner sep=0.75pt]    {$V$};
\draw (582.59,235.03) node [anchor=north west][inner sep=0.75pt]  [font=\normalsize]  {$c_{2}$};
\draw (606.83,291.85) node [anchor=north west][inner sep=0.75pt]  [font=\normalsize]  {$c_{3}$};
\draw (234.5,210.4) node [anchor=north west][inner sep=0.75pt]    {$\beta \rightarrow p_{e}{}_{x}{}_{t}$};
\draw (262.5,87.4) node [anchor=north west][inner sep=0.75pt]    {$e_{i}{}_{j}$};
\draw (281.5,66.4) node [anchor=north west][inner sep=0.75pt]    {$i$};
\draw (298.5,85.4) node [anchor=north west][inner sep=0.75pt]    {$j$};
\draw (385.5,125) node [anchor=north west][inner sep=0.75pt]   [align=left] {$\displaystyle T_{c_{1}}$};
\draw (131.32,114.65) node [anchor=north west][inner sep=0.75pt]  [font=\normalsize]  {$c_{3}$};
\draw (370.5,66.4) node [anchor=north west][inner sep=0.75pt]    {$Poisson$};
\draw (384.5,85.4) node [anchor=north west][inner sep=0.75pt]    {$\lambda _{i}{}_{n}$};
\draw (498.16,96.17) node [anchor=north west][inner sep=0.75pt]    {$j$};
\draw (474.75,95.55) node [anchor=north west][inner sep=0.75pt]    {$V$};
\draw (498.56,59.59) node [anchor=north west][inner sep=0.75pt]    {$i$};
\draw (365.79,297.23) node [anchor=north west][inner sep=0.75pt]  [font=\normalsize] [align=left] {$\displaystyle T_{c_{2}} \cap \ T_{c_{3}}$};
\draw (382.27,261.39) node [anchor=north west][inner sep=0.75pt]    {$\lambda _{e}{}_{x}{}_{t}$};
\draw (329.22,507.4) node [anchor=north west][inner sep=0.75pt]    {${\displaystyle T}$};
\draw (230.5,435.17) node [anchor=north west][inner sep=0.75pt]    {$V$};
\draw (288.98,411.33) node [anchor=north west][inner sep=0.75pt]  [font=\normalsize]  {$c_{1}$};
\draw (348.06,407.89) node [anchor=north west][inner sep=0.75pt]  [font=\normalsize]  {$c_{2}$};
\draw (367.32,456.65) node [anchor=north west][inner sep=0.75pt]  [font=\normalsize]  {$c_{3}$};
\draw (388.06,404.89) node [anchor=north west][inner sep=0.75pt]  [font=\normalsize]  {$c_{*}$};
\draw (245.5,356) node [anchor=north west][inner sep=0.75pt]   [align=left] {\textbf{E. Rewiring noise (}$\displaystyle \eta $\textbf{)}};
\draw (289.06,463.89) node [anchor=north west][inner sep=0.75pt]  [font=\normalsize]  {$c_{*}$};
\draw (368.5,240.4) node [anchor=north west][inner sep=0.75pt]    {$Poisson$};
\draw (297.5,238.4) node [anchor=north west][inner sep=0.75pt]    {$i$};
\draw (316.5,320.23) node [anchor=north west][inner sep=0.75pt]    {$j$};
\draw (300.5,270.4) node [anchor=north west][inner sep=0.75pt]    {$e_{i}{}_{j}$};

\end{tikzpicture}
   \caption{\textbf{Random Mosaic Link Stream Benchmark:} This figure illustrates a five-step process. Step A involves generating a scenario. Step B focuses on removing mosaics with a probability of \(\gamma\) to create an empty community named \(c_*\). Steps C and D add internal and external edges, respectively. Finally, in the last step, each edge in the link stream can be rewired with a probability of \(\eta\).}
    \label{fig: benchmark}
\end{figure}
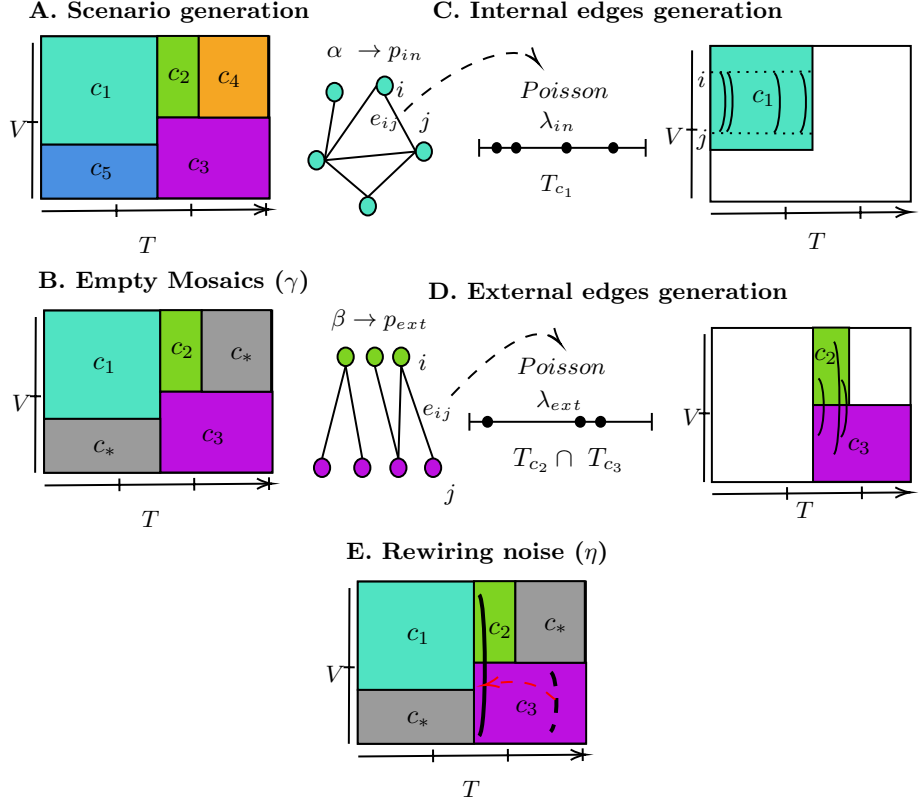

Now, we delve into discussing the Mosaic Benchmark. The proposed framework follows a straightforward five-step process, as depicted in Fig. \ref{fig: benchmark}. The whole procedure is implemented as a user-friendly Python library\footnote[2]{\url{https://pypi.org/project/mosaic-benchmark/}}.
\newpage
We can condense these five steps into two primary stages:
\begin{enumerate}
    \item  Scenario description: The user describes communities using a scenario, either ad-hoc or generated from a provided scenario generator (Step A). Additionally, some communities may be emptied to match real-world properties (Step B).
    
    \item Edge generation: Initially, edges are formed within the communities (Step C), followed by establishing connections between communities with overlapping timeframes (Step D). Some edges may be rewired to introduce imperfections into the community structures (Step E).
\end{enumerate}

\subsection{Scenario description}
\begin{figure}[!ht]
    \centering

\tikzset{every picture/.style={line width=0.75pt}} 

\begin{tikzpicture}[x=0.75pt,y=0.75pt,yscale=-1,xscale=0.9]

\draw   (21.36,20.99) -- (189.79,20.99) -- (189.79,109.98) -- (21.36,109.98) -- cycle ;
\draw  [draw opacity=0][fill={rgb, 255:red, 80; green, 227; blue, 194 }  ,fill opacity=0.3 ] (21.31,20.99) -- (107.12,20.99) -- (107.12,48.25) -- (21.31,48.25) -- cycle ;
\draw  [draw opacity=0][fill={rgb, 255:red, 189; green, 16; blue, 224 }  ,fill opacity=0.3 ] (21.89,82.16) -- (190,82.16) -- (190,109.84) -- (21.89,109.84) -- cycle ;
\draw    (21.1,115.36) -- (184.92,115.21) (75,111.31) -- (75,119.31)(128.9,111.26) -- (128.9,119.26)(182.8,111.21) -- (182.8,119.21) ;
\draw [shift={(186.92,115.2)}, rotate = 179.95] [color={rgb, 255:red, 0; green, 0; blue, 0 }  ][line width=0.75]    (10.93,-3.29) .. controls (6.95,-1.4) and (3.31,-0.3) .. (0,0) .. controls (3.31,0.3) and (6.95,1.4) .. (10.93,3.29)   ;
\draw  [draw opacity=0][fill={rgb, 255:red, 245; green, 166; blue, 35 }  ,fill opacity=0.3 ] (83.4,47.38) -- (189.59,47.38) -- (189.59,82.16) -- (83.4,82.16) -- cycle ;
\draw  [draw opacity=0][fill={rgb, 255:red, 74; green, 144; blue, 226 }  ,fill opacity=0.4 ] (21.31,47.38) -- (83.4,47.38) -- (83.4,83.03) -- (21.31,83.03) -- cycle ;
\draw  [draw opacity=0][fill={rgb, 255:red, 248; green, 231; blue, 28 }  ,fill opacity=0.3 ] (107.12,20.99) -- (190,20.99) -- (190,48.25) -- (107.12,48.25) -- cycle ;
\draw   (240.67,20.49) -- (410,20.49) -- (410,109.62) -- (240.67,109.62) -- cycle ;
\draw  [draw opacity=0][fill={rgb, 255:red, 80; green, 227; blue, 194 }  ,fill opacity=0.3 ] (240.63,20.49) -- (303.04,20.49) -- (303.04,47.79) -- (240.63,47.79) -- cycle ;
\draw  [draw opacity=0][fill={rgb, 255:red, 189; green, 16; blue, 224 }  ,fill opacity=0.3 ] (241.21,82.12) -- (303.04,82.12) -- (303.04,109.48) -- (241.21,109.48) -- cycle ;
\draw    (240.4,114.5) -- (405.12,114.84) (294.31,110.61) -- (294.29,118.61)(348.21,110.73) -- (348.19,118.73)(402.11,110.84) -- (402.09,118.84) ;
\draw [shift={(407.12,114.85)}, rotate = 180.12] [color={rgb, 255:red, 0; green, 0; blue, 0 }  ][line width=0.75]    (10.93,-3.29) .. controls (6.95,-1.4) and (3.31,-0.3) .. (0,0) .. controls (3.31,0.3) and (6.95,1.4) .. (10.93,3.29)   ;
\draw    (237.99,23.42) -- (237.51,109) ;
\draw  [draw opacity=0][fill={rgb, 255:red, 245; green, 166; blue, 35 }  ,fill opacity=0.3 ] (303.04,62.07) -- (327.63,62.07) -- (327.63,81.76) -- (303.04,81.76) -- cycle ;
\draw  [draw opacity=0][fill={rgb, 255:red, 74; green, 144; blue, 226 }  ,fill opacity=0.4 ] (240.63,46.92) -- (303.04,46.92) -- (303.04,82.63) -- (240.63,82.63) -- cycle ;
\draw  [draw opacity=0][fill={rgb, 255:red, 248; green, 231; blue, 28 }  ,fill opacity=0.3 ] (303.04,20.49) -- (327.63,20.49) -- (327.63,62.7) -- (303.04,62.7) -- cycle ;
\draw  [draw opacity=0][fill={rgb, 255:red, 126; green, 211; blue, 33 }  ,fill opacity=0.3 ] (303.04,81.5) -- (327.63,81.5) -- (327.63,109.7) -- (303.04,109.7) -- cycle ;
\draw  [draw opacity=0][fill={rgb, 255:red, 138; green, 12; blue, 12 }  ,fill opacity=0.3 ] (327.63,20.49) -- (409.59,20.49) -- (409.59,37) -- (327.63,37) -- cycle ;
\draw  [draw opacity=0][fill={rgb, 255:red, 245; green, 35; blue, 135 }  ,fill opacity=0.3 ] (327.63,37) -- (409.94,37) -- (409.94,94.03) -- (327.63,94.03) -- cycle ;
\draw  [draw opacity=0][fill={rgb, 255:red, 116; green, 26; blue, 145 }  ,fill opacity=0.3 ] (327.63,94.03) -- (409.59,94.03) -- (409.59,109.07) -- (327.63,109.07) -- cycle ;
\draw    (17.42,22.04) -- (16.95,107.48) ;
\draw   (460.43,20.37) -- (630.05,20.37) -- (630.05,109.96) -- (460.43,109.96) -- cycle ;
\draw  [draw opacity=0][fill={rgb, 255:red, 80; green, 227; blue, 194 }  ,fill opacity=0.3 ] (559.75,20.37) -- (630.05,20.37) -- (630.05,34.59) -- (559.75,34.59) -- cycle ;
\draw    (457.74,23.31) -- (457.26,109.33) ;
\draw  [draw opacity=0][fill={rgb, 255:red, 245; green, 35; blue, 135 }  ,fill opacity=0.3 ] (497.69,47.05) -- (629.64,47.05) -- (629.64,109.41) -- (497.69,109.41) -- cycle ;
\draw [line width=1.5]  [dash pattern={on 1.69pt off 2.76pt}]  (303.04,20.49) -- (303.04,109.7) ;
\draw [line width=1.5]  [dash pattern={on 1.69pt off 2.76pt}]  (327.63,20.49) -- (327.63,109.7) ;
\draw  [draw opacity=0][fill={rgb, 255:red, 65; green, 117; blue, 5 }  ,fill opacity=0.4 ] (506.38,26.31) -- (559.75,26.31) -- (559.75,47.36) -- (506.38,47.36) -- cycle ;
\draw  [draw opacity=0][fill={rgb, 255:red, 248; green, 28; blue, 28 }  ,fill opacity=0.4 ] (482.78,20.37) -- (559.75,20.37) -- (559.75,26.37) -- (482.78,26.37) -- cycle ;
\draw  [draw opacity=0][fill={rgb, 255:red, 144; green, 19; blue, 254 }  ,fill opacity=0.4 ] (460.43,20.37) -- (482.78,20.37) -- (482.78,39.49) -- (460.43,39.49) -- cycle ;
\draw  [draw opacity=0][fill={rgb, 255:red, 80; green, 227; blue, 194 }  ,fill opacity=0.3 ][dash pattern={on 0.84pt off 2.51pt}] (482.92,26.31) -- (506.38,26.31) -- (506.38,39.49) -- (482.92,39.49) -- cycle ;
\draw  [draw opacity=0][fill={rgb, 255:red, 150; green, 254; blue, 19 }  ,fill opacity=0.4 ] (460.43,39.63) -- (506.38,39.63) -- (506.38,47.36) -- (460.43,47.36) -- cycle ;
\draw  [draw opacity=0][fill={rgb, 255:red, 35; green, 133; blue, 245 }  ,fill opacity=0.3 ] (559.75,34.45) -- (616.73,34.45) -- (616.73,47.68) -- (559.75,47.68) -- cycle ;
\draw  [draw opacity=0][fill={rgb, 255:red, 248; green, 231; blue, 28 }  ,fill opacity=0.5 ] (616.73,34.59) -- (630.22,34.59) -- (630.22,47.95) -- (616.73,47.95) -- cycle ;
\draw  [draw opacity=0][fill={rgb, 255:red, 245; green, 166; blue, 35 }  ,fill opacity=0.3 ] (460.74,58.8) -- (468.95,58.8) -- (468.95,109.83) -- (460.74,109.83) -- cycle ;
\draw  [draw opacity=0][fill={rgb, 255:red, 74; green, 144; blue, 226 }  ,fill opacity=0.7 ] (480.68,83.37) -- (497.69,83.37) -- (497.69,109.51) -- (480.68,109.51) -- cycle ;
\draw  [draw opacity=0][fill={rgb, 255:red, 172; green, 30; blue, 188 }  ,fill opacity=0.7 ] (460.74,47.36) -- (497.69,47.36) -- (497.69,58.8) -- (460.74,58.8) -- cycle ;
\draw  [draw opacity=0][fill={rgb, 255:red, 65; green, 117; blue, 5 }  ,fill opacity=0.4 ] (468.95,58.8) -- (497.69,58.8) -- (497.69,83.37) -- (468.95,83.37) -- cycle ;
\draw  [draw opacity=0][fill={rgb, 255:red, 248; green, 231; blue, 28 }  ,fill opacity=0.5 ] (468.95,83.37) -- (474.82,83.37) -- (474.82,96.6) -- (468.95,96.6) -- cycle ;
\draw  [draw opacity=0][fill={rgb, 255:red, 245; green, 166; blue, 35 }  ,fill opacity=0.7 ] (474.82,83.37) -- (480.68,83.37) -- (480.68,102.48) -- (474.82,102.48) -- cycle ;
\draw  [draw opacity=0][fill={rgb, 255:red, 245; green, 35; blue, 245 }  ,fill opacity=0.7 ] (468.95,96.6) -- (474.82,96.6) -- (474.82,109.83) -- (468.95,109.83) -- cycle ;
\draw  [draw opacity=0][fill={rgb, 255:red, 35; green, 213; blue, 245 }  ,fill opacity=0.7 ] (474.82,102.48) -- (480.68,102.48) -- (480.68,109.51) -- (474.82,109.51) -- cycle ;
\draw    (459.73,114.7) -- (624.45,115.05) (513.64,110.82) -- (513.63,118.82)(567.54,110.93) -- (567.53,118.93)(621.44,111.04) -- (621.43,119.04) ;
\draw [shift={(626.45,115.05)}, rotate = 180.12] [color={rgb, 255:red, 0; green, 0; blue, 0 }  ][line width=0.75]    (10.93,-3.29) .. controls (6.95,-1.4) and (3.31,-0.3) .. (0,0) .. controls (3.31,0.3) and (6.95,1.4) .. (10.93,3.29)   ;

\draw (20,7) node [anchor=north west][inner sep=0.75pt]  [font=\small] [align=left] {\textbf{{\small A. Experimental design}}};
\draw (270,7) node [anchor=north west][inner sep=0.75pt]  [font=\small] [align=left] {\textbf{{\small B. Snapshots}}};
\draw (500,7) node [anchor=north west][inner sep=0.75pt]  [font=\small] [align=left] {\textbf{{\small C. Random}}};
\draw (95,115) node [anchor=north west][inner sep=0.75pt]    {$T$};
\draw (315,115) node [anchor=north west][inner sep=0.75pt]    {$T$};
\draw (535,115) node [anchor=north west][inner sep=0.75pt]    {$T$};
\draw (5,60) node [anchor=north west][inner sep=0.75pt]    {$V$};
\draw (225,60) node [anchor=north west][inner sep=0.75pt]    {$V$};
\draw (445,60) node [anchor=north west][inner sep=0.75pt]    {$V$};

\end{tikzpicture}

   \caption{\textbf{Mosaic Partitioning Generation} This figure illustrates three different approaches for generating Mosaic Partitioning in the Benchmark. \textbf{Type A}: User-defined nodes and periods create the desired scenario. \textbf{Type B}: The time domain is divided into multiple frames or snapshots, and node sets are randomly assigned to communities within each frame. \textbf{Type C}: Communities' time interval and node-set sizes are distributed inhomogeneously, covering the entire link stream.}
    \label{fig: Scanario_description}
\end{figure}
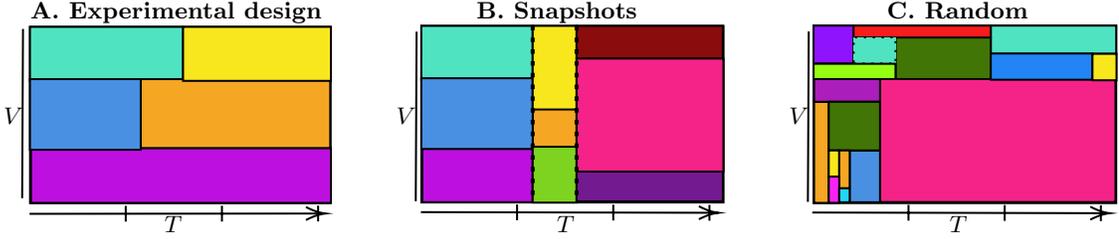

In the proposed framework, we use Mosaic partitioning to generate modular link streams according to a scenario. Mosaic partitioning consists of multiple communities in which nodes interact within and between them.  It can be generated using three proposed scenario generators in our Python library: Experimental, Snapshots, and Random. A visualization can be found in Fig. \ref{fig: Scanario_description}.

\subsubsection{Experimental:}
Using an experimental setup, we can intentionally generate various Mosaic communities with specific node counts and time intervals. This enables us to evaluate the performance and precision of a community detection algorithm under straightforward scenarios, gradually increasing complexity to gain insights into the algorithm's capabilities and constraints.

\subsubsection{Snapshots:}
This scenario generator simulates low-resolution temporal networks resembling snapshots. It partitions the time domain into \(k\) segments defined by either fixed or variable window sizes. In the fixed case, each segment has a size of \(\frac{|T|}{k}\), while in the varying case, the time domain is randomly divided. Then, we establish static communities for each segment, ensuring they consist of at least two or more nodes.
\begin{algorithm}
    \caption{Snapshot Scenario Generator}
    \label{Algo:Snapshot}
    \begin{algorithmic}[1]
        \Procedure{Snapshot}{$T, V, k, \text{Mode}$} 
            \State Create an empty list $\mathcal{C}$
            \State Divide the time domain into $k$ segments with fixed or varying window sizes and name it $S$
            \For{each $s$ in $S$}
                \State Distribute the nodes \(V\) randomly into static communities and name the collection $U$
                \For{each $u$ in $U$}
                    \State Add $(u, s)$ to $\mathcal{C}$
                \EndFor
            \EndFor
            \State \textbf{return} $\mathcal{C}$
        \EndProcedure
    \end{algorithmic}
\end{algorithm}

\subsubsection{Random:}
\label{random scenario}
Generating random scenarios leading to mosaics with different node counts and time intervals is essential for comprehending the capabilities and constraints of a dynamic community detection algorithm. To accomplish this objective, we will suggest a simple procedure for constructing random communities using \(k\) iterations in a recursive algorithm.
\begin{algorithm}
    \caption{Random Scenario Generator}
    \label{Algo: random}
    \begin{algorithmic}[1]
        \Procedure{Random}{$T, V, k$}
            \State Create a Mosaic $M=(T, V)$
            \For{$i$ in $0$ to $k$}
                \If{$M$.sub-mosaics is empty}
                    \State Divide $M$ into 4 sub-mosaics and store them in $M$.sub-mosaics
                \Else
                    \State Select one sub-mosaics from $M$.sub-mosaics and divide it into 4 sub-mosaics
                \EndIf
            \EndFor
            \State Flatten the Mosaic \(M\) into $\mathcal{C}$
            \State \textbf{return} $\mathcal{C}$
        \EndProcedure
    \end{algorithmic}
\end{algorithm}
Once we have obtained the algorithm's output, we will proceed to remove communities consisting of just one node or communities that exist for less than a predefined time interval.

\subsubsection{Emptying Mosaics.} We consider that Mosaics are assigned to an empty Mosaic \(c_*\) with a probability of \(\gamma\). We mean that within this empty Mosaic, no edges can be active that originate from either inside or outside, affecting the nodes contained within it.
\subsection{Generating edges}

\label{Generating edges}
This part focuses on generating edges between nodes within and across different communities. We will follow two steps: creating a Backbone connectivity network to establish static connections and using the Poisson Point Process to add a temporal dimension above it; refer to Algorithm \ref{Algo: EdgeGeneration}.
\begin{algorithm}
\caption{Edges Generation}
\label{Algo: EdgeGeneration}
\begin{algorithmic}[1]
\Procedure{EdgesGeneration}{$\mathcal{C}, \alpha ,\lambda,\beta $} 
    \State \text{Create an empty list \(E\)}
    \For{$c$ in $\mathcal{C} \backslash c^*$} \Comment{\textcolor{blue}{Generate internal edges}}
    \State \(p^c_{in} = (|V_c|-1)^{\alpha-1}\)
    \State List \(S\) =\textbf{BackboneConnectivity}(\(c\),\(c\),\(p^c_{in}\))
     \For{$e$ in $S$}
    \State Add \textbf{PoissonProcessEdge}(\(e\),\(P_c\),\(\lambda^{cc}\)) to \(E\)
    \EndFor
    \EndFor
    \For{$(c, c^\prime)$ in $\binom{\mathcal{C} \backslash c^*}{2}$}\Comment{\textcolor{blue}{Generate external edges}}
    \State \(p^{cc^\prime}_{ext}=\beta ((|V_c|+|V_c'|)-1)^{\alpha-1} \)
    \State List \(S\) =\textbf{BackboneConnectivity}(\(c\),\(c^\prime\),\(p^{cc^\prime}_{ext}\))
      \For{$e$ in $S$}
    \State Add \textbf{PoissonProcessEdge}(\(e\),\(P_c \cap P_{c'}\),\(\lambda^{cc^\prime}\)) to \(E\)
    \EndFor
    \EndFor
     \State \textbf{return} \(E\)
\EndProcedure
\end{algorithmic}
\end{algorithm}
\subsubsection{Backbone connectivity network.}
\label{Backbone connectivity network}
This Benchmark assumes that the connectivity between nodes, whether through internal or external edges, remains stable throughout the specified period. This is why we refer to it as the backbone connectivity network. A backbone connectivity network with a parameter \(p\) is a random graph in which each edge is present with probability \(p\), independent of others. 

We would like to emphasize that for establishing a well-defined internal structure of a community, it is necessary to utilize an appropriate range of values for \(p\). This range's selection should depend on the number of vertices within the community. To achieve this, we will adopt the model described in \cite{cazabet2020evaluating}, which provides the formula for \(p^c_{in}\) as follows:

\[p^c_{in} = (|V_c|-1)^{\alpha-1}\]

Here, \(\alpha \in (0, 1]\) is a hyperparameter named community density coefficient shared between communities. When the value of \(\alpha\) is increased, the probability of \(p^c_{in}\) also increases, leading to denser clusters. If \(\alpha\) is set to 1, each community in Mosaic becomes a clique.

The external probability between two communities \(c\) and \(c'\) denoted as \(p^{cc'}_{ext}\) is defined as:
\[p^{cc'}_{ext} = \beta (|V_c|+|V_c'|-1)^{\alpha-1}\]
This hyperparameter \(\beta \in [0,1]\) is related to ``community identifiability," which is shared among all communities. Increasing the value of \(\beta\) results in more external edges between communities, making it more challenging for algorithms to identify each community as a separate cluster. In other words, \(\beta\) controls the external density of backbone connectivity by treating two communities as a single entity.

\subsubsection{Poisson Point Process.}
\label{Poisson Point Process}
To simplify the analysis, we assumed that the edges in a given backbone connectivity network follow a memory-less stochastic process for their activation times. For each edge \(e=(i, j )\) in the backbone connectivity network, we generate an independent and identically distributed random Poisson point process with a rate parameter \(|T|\lambda\). This rate parameter determines the average number of this edge active times within the time frame \(T\). Then, we use the uniform distribution to distribute this number of occurrences in the selected period. This means the edge time arrivals are uniformly spread over the interval \(T\) \cite{ross2014introduction}.
To establish connections within a community, we set \(T\) equal to \(T_c\) and \(\lambda =\lambda^c_{in}\). Conversely, when creating the connections between two different communities \(c\) and \(c^\prime\), \(T\) is defined by the overlap between \(T_c\) and \(T_{c^\prime}\). Furthermore, to generate external edges between communities \(c\) and \(c'\), we utilize a coefficient \(\lambda^{cc'}_{ext}\). 

Combining these, to create the temporal dimension, we need a symmetric matrix \(\lambda\) of size \(k\times k\), where \(k\) represents the number of communities. The main diagonal of this matrix will be utilized for generating internal edges, and non-diagonal elements can be employed for external edges if there is a non-empty time overlap (\(T_c \cap T_c' \neq \emptyset\)) between the communities \(c\) and \(c'\).

In both the step of generating backbone connectivity networks and adding a Poisson point process layer, each community is handled independently, which can be efficiently parallelized. This enables handling large networks in a reasonable time.
Additionally, finding an upper bound for memory and time complexity can not be provided due to dependence both on time and structure. 

\subsubsection{Rewiring noise.}
A prior study \cite{kobayashi2019structured} proposed that it is possible to distinguish a stable core within communities from random, short-lived fluctuations in real temporal networks. In our Benchmark, edges go through a rewiring process with a probability of \(\eta=[0,1]\) to highlight imperfections in community structures. During this step, for edge \((u,v,t)\) selected for rewiring, we randomly choose two communities \(c\neq c_*\) and \(c'\neq c_*\) with non-empty time intersection. We then select two nodes \(u \in V_{c}\) and \(v \in V_{c'}\) where \(u\neq v\), with a timestamp \(t\) randomly chosen within the \(T_{c}\cap T_{c'} \) time frame.
\section{Experiments}
In this section, we use different community detection algorithms on an instance of our framework to assess their performance in identifying communities compared to the ground truth. Among the many dynamic community detection algorithms available, we have selected four that operate by aggregating the link stream into snapshots using window sizes. These algorithms are previously implemented in ``tnetwork" Python package \cite{cazabet2023tnetwork}, as detailed in the existing literature review \cite{rossetti2018community}.

The algorithms compared in this paper are the following:
\begin{itemize}
    \item \textbf{No-Smoothing:} the approach involves applying a static algorithm (in this case, the Louvain method) to each snapshot. Then, the most similar communities in consecutive steps are matched based on the Jaccard Coefficient with a coefficient named \(\theta\), set to 0.3 here.
    
    \item \textbf{Implicit-Global:} in this method, the Louvain algorithm is executed at each snapshot, but instead of initializing it with each node in its own community, the previous partition is utilized as the seed.
    
    \item \textbf{Label Smoothing:} this method first identifies communities in each slice. Then, attempts are made to match communities across different snapshots, forming a survival graph. A community detection algorithm is applied to this survival graph, resulting in dynamic snapshot communities.
    
    \item \textbf{Smoothed-Graph:} in this approach, the Louvain method is run at each slice \(t\) on a graph with a smoothed adjacency matrix defined as follows:
    \[
    \text{A}^t_{ij} = \alpha \cdot \text{A}^t_{ij} + (1 - \alpha) \cdot C^{t-1}_{ij},
    \]
    where \(C^{t-1}_{ij} = 1\) if nodes \(i\) and \(j\) belong to the same community at step \(t - 1\), and 0 otherwise.
\end{itemize}
We considered a link stream \(L\) consisting of 100 nodes \(|V|=100\) interacting over the time interval \(T=[0,100)\). Then, we created a scenario using a Random scenario generator(as detailed in \ref{random scenario}) with an iteration parameter \(k=30\). Additionally, we applied a probability \(\mu=0.2\) to empty Mosaic communities, resulting in \(|\mathcal{C}|=61\) communities with an average node count of \(\bar{V}_c=11.7\) and an average time interval of \(\bar{T}_c=10.2\), excluding the empty community \(c_*\).

For the edge generation phase, we fine-tuned the parameters \(\alpha=0.9\) and \(\beta=0.1\). A high \(\alpha\) value signifies strong connections among nodes within the internal backbone connectivity networks, while a low \(\beta\) value indicates sparse connections between communities in the external backbone connectivity networks. This configuration results in clear and distinguishable communities regarding the backbone structure.

To further emphasize this characteristic, we maintained the same values for \(\lambda_{in}=0.4\) and \(\lambda_{ext}=0.1\) for all communities, signifying that, in the Poisson Point Process, the likelihood of an edge forming within a community is higher than it is for an external one. It resulted in 28365 edges. Additionally, we exclude the rewiring process for temporal edges. 

Since the chosen dynamic community detection methods are not tailored for link stream cases and are designed for low-resolution temporal networks, we convert our sample modular link stream to snapshots with a relatively small fixed window size 2. Subsequently, we applied community detection algorithms to the aggregated snapshots and extracted these communities for the purpose of comparison and visualization.

\begin{figure}[!ht]
    \centering
    \begin{subfigure}{0.4\textwidth}
        \includegraphics[keepaspectratio, height=5cm]{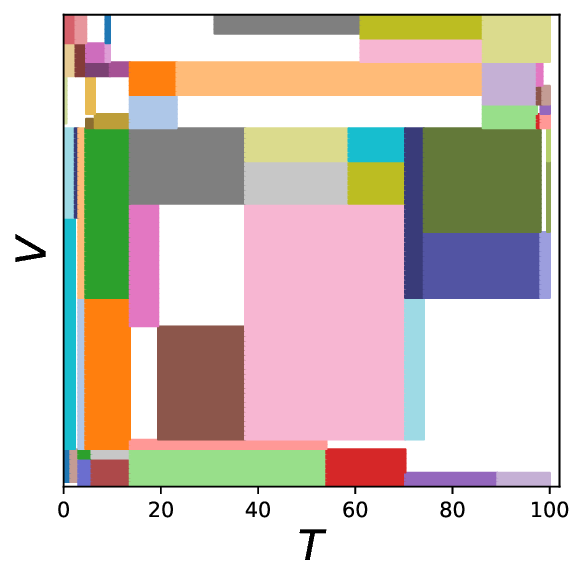}
        \caption{Ground Truth}
    \end{subfigure}
    \hfill
    \begin{subfigure}{0.6\textwidth}
        \begin{subfigure}{0.5\linewidth}
            \includegraphics[keepaspectratio, height=4cm]{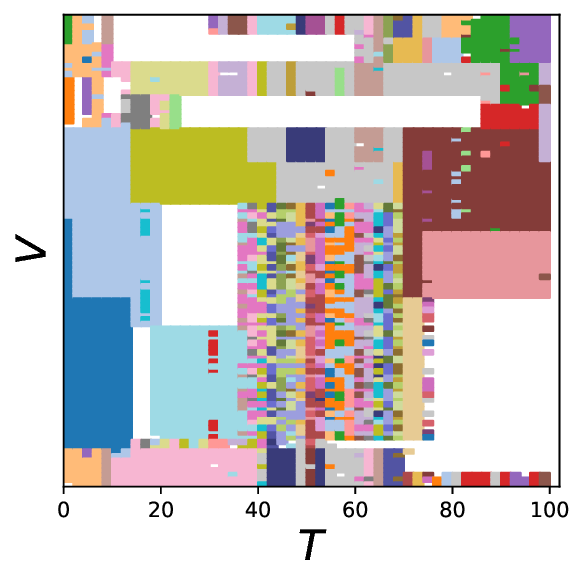}
            \caption{No-Smooting}
        \end{subfigure}
        \begin{subfigure}{0.5\linewidth}
            \includegraphics[keepaspectratio, height=4cm]{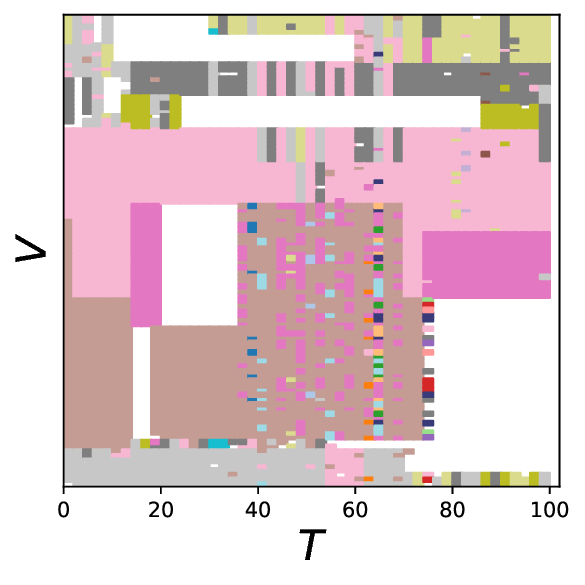}
            \caption{Label smoothing}
        \end{subfigure}
        \begin{subfigure}{0.5\linewidth}
            \includegraphics[keepaspectratio, height=4cm]{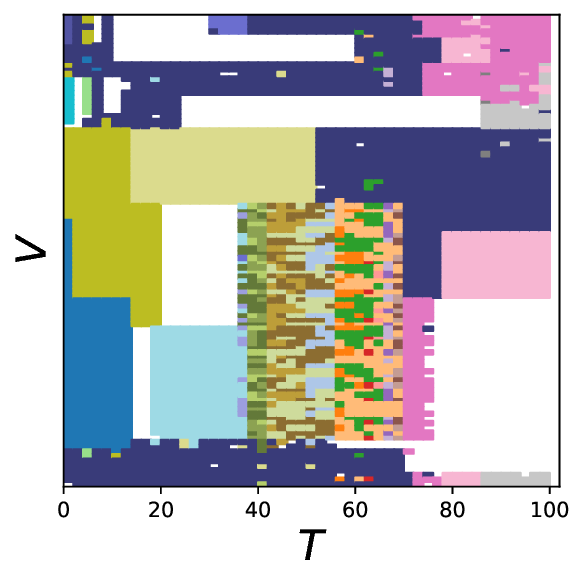}
            \caption{Implicit global}
        \end{subfigure}
        \begin{subfigure}{0.5\linewidth}
            \includegraphics[keepaspectratio, height=4cm]{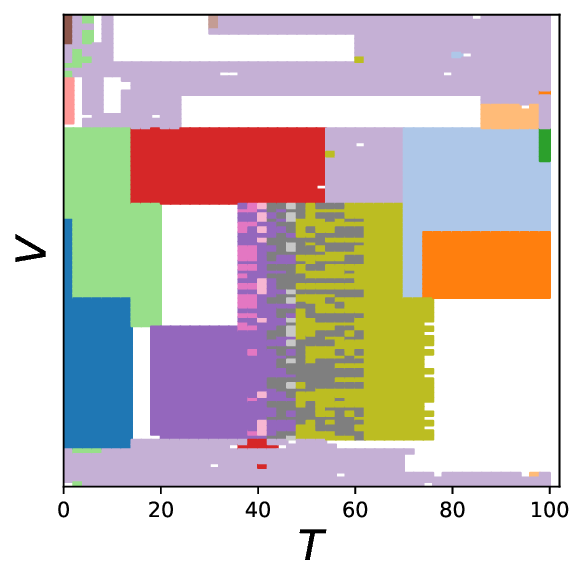}
            \caption{Smoothed graph}
        \end{subfigure}
   
    \end{subfigure}
   \caption{\textbf{Experiments:} Comparison of partitions obtained using all methods on a sample Random scenario}
    \label{fig: Analysis}
\end{figure}

As depicted in Fig. \ref{fig: Analysis}, it is evident that different algorithms can yield varying interpretations of a benchmark network instance. Assessing the accuracy of community detection algorithms under such circumstances can pose a significant challenge. Similarly, when employing aggregation techniques on a link stream, diverse interpretations may arise. In certain social and political contexts, these variations can be contentious and potentially lead to misleading conclusions.

To conduct a more rigorous experiment, we introduce a new parameter denoted as \(\phi=1-\alpha=\beta\). A higher \(\phi\) parameter value results in communities being less identifiable. We varied the \(\phi\) value from 0 to 0.5 with the step of 0.1, did the same generation process, and aggregated the link streams to snapshots using a window size of 2, then identified communities ten times with the algorithms.
\begin{figure}[!ht]
    \centering
    \includegraphics[width=14cm]{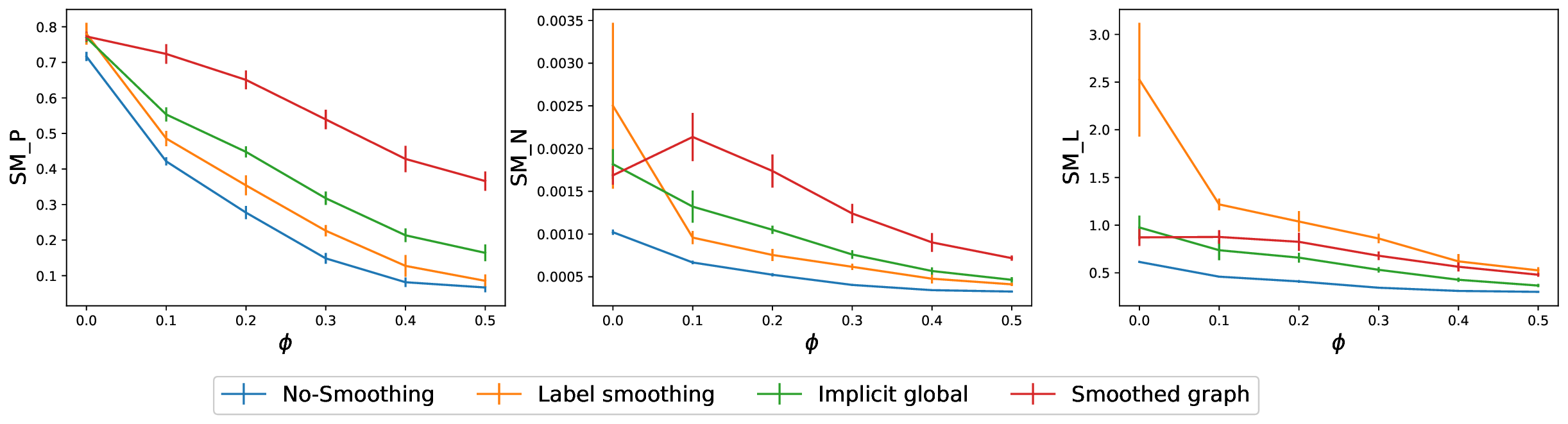}
    \caption{\textbf{Experiments:} Comparison of smoothness measures obtained using all methods on a sample of a Random scenario.}
    \label{fig: Analysis2}
\end{figure}

Smoothness values were determined utilizing the formulas for smoothness presented in the work by Cazabet et al. \cite{cazabet2020evaluating}, namely SM-P, SM-N, and SM-L. In all these smoothness metrics, a higher value indicates superior performance. The obtained values were averaged for each \(\phi\).

As depicted in Figure \ref{fig: Analysis2}, in terms of smoothness, two methods have high scores for the three aspects: Implicit-Global and Smoothed-Graph. Label-smoothing has the highest scores in most settings for the SM-L scores, which measure
label smoothness. No Smoothing is the least stable in most cases. These results support the findings of other benchmarks presented in \cite{cazabet2020evaluating}.  
\section{Discussion and Conclusions}
In summary, we introduced the Mosaic benchmark networks as a new framework for generating modular link streams. These temporal networks enable the evaluation of dynamic community detection algorithms in terms of accuracy and performance. Additionally, we can provide a quantitative assessment based on community definition, stability, and scalability by simulating adaptable ground truth to determine an algorithm's capabilities and constraints. Furthermore, our framework acts as a foundational platform for assessing algorithms tailored for link streams, especially in the later stages of research in this domain.

The time complexity of this framework can benefit from parallelization because it avoids SBM calculations for each small step, and the future does not depend on past or present data. Moreover, when simulating link streams with millions of edges, it is crucial to efficiently organize edge and node storage. Multi-layered hash functions can enhance memory allocation for storing network edges instead of a large edge stream. To create directed modular link streams, we propose using the benchmark network with an asymmetric Poisson rate matrix (\(\lambda\)), along with two additional parameters, \(\alpha_{out}\) and \(\beta_{out}\) for edge generation phase. However, it is essential to carefully check if created communities match the definition of communities in real-world directed link streams.

We applied our framework to explore differences among communities identified by various dynamic community detection algorithms. We emphasize that the aggregation method using window sizes has its limitations, and there is a crucial need for a more comprehensive investigation into effectively handling continuously timed edges rather than aggregating them. 

In conclusion, it is important to explore the characteristics of communities formed through the provided scenario generators. This framework incorporates various parameters such as \(\gamma\) for emptying communities, and (\(\alpha\), \(\beta\), \(\lambda\)) for edge generation. Furthermore, the parameter \(\eta\) plays a crucial role in evaluating an algorithm's robustness. We advocate for further research to enhance the comprehension of dynamic community algorithms, achieved through the fine-tuning of these parameters.

\section*{Funding} We thank the University of Zurich and the Digital Society Initiative for (partially) financing this project conducted by Yasaman Asgari. The work has been supported by the ANR grant DARLING ANR-19-CE48-0002, the ANR grant BITUNAM ANR-18-CE23-0004, and the CHIST-ERA grant CHIST-ERA-19-XAI-006, for the GRAPHNEX ANR-21-CHR4-0009 project.
We also thank Victor Brabant for his valuable discussions.
\bibliographystyle{ieeetr}
\bibliography{main}
\end{document}